\documentclass[twocolumn]{aastex631}

\usepackage{amssymb,amsmath,mathtools,xfrac}
\usepackage{multirow}
\usepackage{tikz}
\usepackage{makecell}
\usepackage{acronym}
\usetikzlibrary{bayesnet}
\usepackage{comment}

\acrodef{sGRB}[sGRB]{short gamma-ray burst}
\acrodef{GW}[GW]{gravitational wave}
\acrodef{EM}[EM]{electromagnetic}
\acrodef{BNS}[BNS]{binary neutron star}

\begin{document}

\title{Unpacking merger jets: a Bayesian analysis of GW170817, GW190425\\ and electromagnetic observations of short gamma-ray bursts}

\author[0000-0001-7628-3826]{Fergus Hayes}
\email{fergus.hayes@glasgow.ac.uk}
\affiliation{SUPA, School of Physics and Astronomy, University of Glasgow, Glasgow, G12 8QQ, UK}

\author[0000-0002-1977-0019]{Ik Siong Heng}
\affiliation{SUPA, School of Physics and Astronomy, University of Glasgow, Glasgow, G12 8QQ, UK}

\author[0000-0001-5169-4143]{Gavin Lamb}
\affiliation{Astrophysics Research Institute, Liverpool John Moores University, IC2 Liverpool Science Park, 146 Brownlow Hill, Liverpool, L3 5RF, UK}
\affiliation{School of Physics and Astronomy, University of Leicester, University Road, Leicester, LE1 7RH, UK}

\author[0000-0002-0030-8051]{En-Tzu Lin}
\affiliation{Institute of Astronomy, National Tsing Hua University, Hsinchu 30013, Taiwan}

\author[0000-0002-6508-0713]{John Veitch}
\affiliation{SUPA, School of Physics and Astronomy, University of Glasgow, Glasgow, G12 8QQ, UK}

\author[0000-0003-2198-2974]{Michael J. Williams}
\affiliation{SUPA, School of Physics and Astronomy, University of Glasgow, Glasgow, G12 8QQ, UK}



\begin{abstract}
    We present a novel fully Bayesian analysis to constrain short gamma-ray burst jet structures associated with cocoon, wide-angle and simple top-hat jet models, as well as the binary neutron star merger rate.
    These constraints are made given the distance and inclination information from GW170817, observed flux of GRB\,170817A, observed rate of short gamma-ray bursts detected by Swift, and the neutron star merger rate inferred from LIGO's first and second observing runs.
    A separate analysis is conducted where a fitted short gamma-ray burst luminosity function is included to provide further constraints.
     The jet structure models are further constrained using the observation of GW190425 and we find that the assumption that it produced a GRB\,170817-like short gamma-ray burst that went undetected due to the jet geometry is consistent with previous observations.
    We find and quantify evidence for low luminosity and wide-angled jet structuring in the short gamma-ray burst population, independently from afterglow observations, with log Bayes factors of $0.45{-}0.55$ for such models when compared to a classical top-hat jet.
    Slight evidence is found for a Gaussian jet structure model over all others when the fitted luminosity function is provided, producing log Bayes factors of $0.25{-}0.9\pm0.05$ when compared to the other models.
    However without considering GW190425 or the fitted luminosity function, the evidence favours a cocoon-like model with log Bayes factors of $0.14\pm0.05$ over the Gaussian jet structure.
    We provide new constraints to the binary neutron star merger rates of $1{-}1300$\,Gpc$^{-3}$\,yr$^{-1}$ or $2{-}680$\,Gpc$^{-3}$\,yr$^{-1}$ when a fitted luminosity function is assumed.
\end{abstract}



\section{Introduction}

The joint detection of both \ac{GW} GW170817~\citep{abbott2017gw170817}, and counterpart \ac{sGRB} GRB\,170817A~\citep{goldstein2017ordinary,savchenko2017integral}, followed by the detection of kilonova AT\,2017gfo~\citep{mccully2017rapid,evans2017swift} not only solidified the belief that \ac{sGRB}s are produced from the merger of \ac{BNS} systems, but also began the era of \ac{GW} multimessenger astronomy~\citep{abbott2017multi}.
The combination of both the \acrodef{GW}[GW]{gravitational wave}
 and \ac{GW} data gave insight into various problems that a detection through a single data channel could not provide; ranging from cosmology~\citep{ligo2017hubble}, the origin of the abundance of heavy elements in the universe~\citep{tanvir2017emergence}, tests of general relativity and the speed of gravity~\citep{abbott2017gravitational} among others.

However the detection of the event not only provided answers but also provoked questions, as it was inferred through \ac{GW} parameter inference that the event was exceptionally nearby at only $40$\,Mpc~\citep{abbott2019properties}, giving an observed isotropic luminosity of the event of $10^{47}$\,erg\,s$^{-1}$, three orders of magnitude lower than that observed for any other \ac{sGRB}~\citep{abbott2017gravitational}.
It was also inferred through \ac{GW} parameter inference that the event was viewed at a wide angle of $14-41^{\circ}$ from the central axis~\citep{abbott2019properties}.
This lead to the hypothesis that the jet of GRB\,170817A exhibited some wide-angle structure to produce the observed flux, and that it may still have had a typically luminous central jet component.
The long duration observations of the event's afterglow across the \ac{EM} spectrum provided further evidence for this claim~\citep{troja2017x,margutti2018binary,lyman2018optical,d2018evolution,mooley2018mildly,ruan2018brightening,troja2018outflow,alexander2018decline,troja2020thousand}.

While evidence for this wide-angled structure is provided by the multitude of afterglow observations, the functional form of the luminosity profile over viewing angle remains in question \citep[e.g.][]{granot2017, duffell2018, gottlieb2018, lamb2018grb, mooley2018mildly, troja2018outflow, ioka2019spectral, beniamini2019, fraija2019, lamb2019, salafia2019, biscoveanu2020constraining, lamb2020, takahashi2021}.
Accurate modelling of this jet structure is important in both understanding the astrophysics of the event, and in preventing systematic biases in any multimessenger analysis that must make assumptions about the jet geometry \citep{nakar2021afterglow, lamb2021U}, which include constraints on the Hubble constant~\citep{ligo2017hubble} and the \ac{BNS} merger rate~\citep{wanderman2015rate}.

With the promise of future \ac{GW} \ac{BNS} merger detections~\citep{abbott2020prospects}, a jet structure model that best represents the data of joint detection events should be discerned~\citep{hayes2020comparing}.
However, current analyses are limited to the data provided by GW170817/GRB\,170817A (including AT\,2017gfo), \ac{GW} \ac{BNS} merger event GW190425~\citep{abbott2020gw190425}, as well as the population of \ac{sGRB}s detected independently of \ac{GW} detection~\citep{poolakkil2021fermi,lien2016third}.
Previous similar work has considered constraining jet structure models with joint \ac{GW} and \ac{EM} detections using a Bayesian analysis~\citep{biscoveanu2020constraining,hayes2020comparing,farah2020counting}.
The historical rate of detected \ac{sGRB} and \ac{GW}s has also been used in Bayesian analyses to constrain the jet structure~\citep{williams2018constraints,sarin2022linking}.
Work by \citet{mogushi2019jet} and \citet{tan2020jet} combine the detection rates of \ac{sGRB} and \ac{GW}s with the prompt emission data of GRB\,170817A and the parameter inference results of GW170817, along with assuming a luminosity function fitted by short gamma-ray burst events with known redshift.
The jet structure has been constrained given the weak gamma-ray emission detected by the \textit{INTEGRAL} detector coincident with the gravitational wave event GW190425 in the work by \citet{saleem2020energetics}.
In this work, we put forward a comprehensive Bayesian framework that combines the prompt emission data from GRB\,170817A, \ac{GW} parameter inference posteriors of GW170817 and GW190425, \ac{GW} informed \ac{BNS} merger rate, as well as the detection rate of \ac{sGRB}s by the Neil Gehrels Swift Observatory (Swift) detector.
This analysis is then further combined with the information provided by a luminosity function fitted by \ac{sGRB} events with known redshifts to provide tighter constraints, with the caveat of also introducing bias into the analysis.
We provide parameter constraints and model comparison results between a classical top-hat jet structure and three different jet structures with wide-angled structuring: a Gaussian, power-law and double Gaussian jet.
These results are presented alongside constraints on the intrinsic luminosity (when it is not fitted by the luminosity function), and the merger rate for both cases.

The physical model assumed is detailed in Section~\ref{sec:BG}, before the analysis method and data are laid out in Section~\ref{sec:anal}. In Section~\ref{sec:results} we report the results of the analysis, both for when the fitted luminosity function is incorporated and when it is not.
In Section~\ref{sec:disc} the implications of the results are discussed and a conclusion provided in Section~\ref{sec:conc}.

\section{Background}\label{sec:BG}

The \ac{sGRB} data consists of the observed T90 integrated flux $F$ as well as the number of observed \ac{sGRB}s $N_{\text{EM}}$.
The average T90 integrated flux $\hat{F}$ is related to the isotropic equivalent luminosity $L_{\text{iso}}$ at a given viewing angle $\theta_{v}$, as well as a redshift $z$ dependent luminosity distance $d_{L}$ and $k$-correction $k$:
\begin{equation}\label{equ:flux}
\hat{F} = \frac{L_{\text{iso}}(\theta_{\text{v}})}{4\pi d_{L}(z)^{2}k(z)}.
\end{equation}

The mean number of \ac{sGRB}s $\hat{N}_{\text{EM}}$ observed by a detector within a duration $T$ that covers an area of sky equal to $\Delta\Omega$ depends on the redshift, viewing angle and intrinsic luminosity $\Lambda=\{z,\theta_v,L_0\}$ through the number density:
\begin{equation}\label{equ:EMN}
\hat{N}_{\text{EM}} = \int \frac{d\hat{N}_{\text{EM}}}{d\Lambda}\left[\int p(D_{\text{EM}}|\Lambda)dD_{\text{EM}}\right]d\Lambda,
\end{equation}
where $D_{\text{EM}}$ denotes the selection effects of the detector, such that $D_{\text{EM}}=1$ if a detection is made and $D_{\text{EM}}=0$ otherwise.

We consider \ac{sGRB}s detected by the Swift instrument, which has a detector response determined empirically in \citep{lien2014probing} to fit:
\begin{equation}
\int p(D_{\text{EM}}|\Lambda)dD_{\text{EM}} = \begin{cases}
0 & \text{if }\hat{F}<F_{\text{thr}}, \\
a\frac{b + c\hat{F}/F_{0}}{1 + \hat{F}/dF_{0}} & \text{else,}
\end{cases}
\end{equation}
where $a=0.47$, $b=-0.05$, $c=1.46$, $d=1.45$, $F_{0}=6\times10^{-6}$\,erg\,s$^{-1}$\,cm$^{-2}$ and $F_{\text{thr}}=5.5\times10^{-9}$\,erg\,s$^{-1}$\,cm$^{-2}$.

The relation between the number density and the physical parameters of $\Lambda$ is:
\begin{equation}\label{equ:Nev}
\frac{d\hat{N}_{\text{EM}}}{d\Lambda} = T\frac{\Delta\Omega}{8\pi}\frac{\mathcal{R}_{\text{GRB}}(z)}{1+z}\frac{dV(z)}{dz}p(L_{0}|\Sigma)\sin\theta_{v},
\end{equation}
where $V(z)$ is the co-moving volume, $\mathcal{R}_{\text{GRB}}$ is the rate of \ac{sGRB}s and $p(L_0|\Sigma)$ is the intrinsic luminosity function given the hyperparameter $\Sigma$.

\subsection{Short gamma-ray burst rate}

The rate of \ac{sGRB}s is assumed to be in the form:
\begin{equation}\label{equ:localsGRB}
\mathcal{R}_{\text{GRB}}(z) = R_{\text{BNS}}R_{\text{GRB}}(z),
\end{equation}
where $R_{\text{BNS}}$ is the local rate of \ac{BNS} mergers and $R_{\text{GRB}}(z)$ is defined so that $R_{\text{GRB}}(0)=1$.
This assumes that every \ac{BNS} merger results in a \ac{sGRB}, and that the number of \ac{sGRB}s produced by neutron star-black hole mergers is negligible.

The form of $R_{\text{GRB}}(z)$ can be assumed to follow the star formation rate $R_{\ast}(z)$ convolved with the probability distribution of the delay time between the system formation and the eventual merger that leads to the \ac{sGRB} $ P(t) $~\citep{wanderman2015rate}:
\begin{equation}\label{equ:Rstar}
R_{\text{GRB}}(z) \propto \int_{t_{\text{min}}}^{T(\infty)-T(z)} \frac{R_{\ast}(z_{\ast})}{1+z_{\ast}}P(t)dt,
\end{equation}
where $z_{\ast} = z(T(z)+t)$ is the redshift when the system was formed, $T(z)$ is the \textit{look-back time} and $t_{\text{min}}$ is the minimum delay time.
This minimum delay time is set to $20\,$Myr and $P(t)\propto 1/t$ according to~\citep{guetta2006batse}.
The star formation rate is assumed to be of the form~\citep{cole20012df}:
\begin{equation}
R_{\ast}(z) \propto \frac{a+bz}{1+(z/c)^{d}}H(z),
\end{equation}
where the parameter values are taken from \citep{hopkins2006normalization} to be $a=0.017$, $b=0.13$, $c=3.3$ and $d=5.3$.

\subsection{Cosmology}

A flat, vacuum dominated universe is assumed.
The co-moving volume distribution over redshift is defined:
\begin{equation}\label{equ:CMV}
\frac{dV(z)}{dz} = 4\pi \frac{c}{H(z)}\left(\frac{d_{L}(z)}{1+z}\right)^{2}.
\end{equation}
For a flat cosmology, the luminosity distance is related to the redshift by:
\begin{equation}\label{equ:dL}
d_{L}(z) = (1+z)\frac{c}{H_{0}}\int_{0}^{z}\frac{H_0}{H(z')}dz',
\end{equation}
where $H_{0}$ is the Hubble constant and $H(z)$ is equal to:
\begin{equation}
H(z) = H_0\sqrt{\Omega_{m}(1+z)^{3} + \Omega_{\Lambda}}.
\end{equation}
Here $\Omega_{m}=0.308$ and $\Omega_{\Lambda}=0.692$ are the matter density and dark energy density respectively~\citep{hogg1999distance}, with values taken from \citep{adam2016planck} along with $H_0=67.8\,$km\,s$^{-1}$\,Mpc$^{-1}$.

The look-back time, defined as the time between when a source emits light at redshift $z$ and the time it is detected, is then:
\begin{equation}\label{equ:Tz}
T(z) = \frac{1}{H_0}\int_{0}^{z}\frac{H_0}{(1+z')H(z')}dz'.
\end{equation}
For a flat, vacuum dominated universe, the inverse function has an analytical expression~\citep{petrillo2013compact}: 
\begin{equation}
z(T) = \left(\frac{\Omega_{\Lambda}}{\Omega_{m}}\right)^{1/3}\left[\left(\frac{1+W(T)}{1-W(T)}\right)^{2} - 1 \right]^{1/3} - 1,
\end{equation}
where:
\begin{equation}
W(T) = \exp\left[\ln\left(\frac{1+\sqrt{\Omega_{\Lambda}}}{1-\sqrt{\Omega_{\Lambda}}}\right) - 3H_0 \sqrt{\Omega_{\Lambda}}T\right].
\end{equation}

The $k$-correction accounts for the cosmological redshifting in the intrinsic \ac{sGRB} spectrum with respect to the detector's spectrum~\citep{bloom2001prompt}:
\begin{equation}\label{equ:kcor}
k(z) = \frac{\int_{\nu_{s,1}/(1+z)}^{\nu_{s,2}/(1+z)}\nu f(\nu)d\nu}{\int_{\nu_{1}}^{\nu_{2}}\nu f(\nu)d\nu}.
\end{equation}
Here we assume the form of $f(\nu)$ follows the Band function described in \citep{band1993batse}.

\subsection{Intrinsic and isotropic equivalent luminosity}

The luminosity structure of a gamma-ray burst is defined to be:
\begin{equation}\label{equ:intrinsicE}
L(\theta) = L_{0}y_{L}(\theta),
\end{equation}
where $ L_{0} $ is the intrinsic luminosity at $\theta=0$.
It is assumed that the distribution of intrinsic luminosity $L_0$ follows a Schechter function:
\begin{equation}\label{equ:pLsig}
p(L_{0}|\Sigma=\{L_{0}^{\ast},\gamma\}) \propto \left(\frac{L_{0}}{L_{0}^{\ast}}\right)^{-\gamma}e^{-L_{0}/L_{0}^{\ast}},
\end{equation}
for $L_{\text{min}}\le L_0 \le L_{\text{max}}$ where $L_{\text{min}}=10^{-3}L_0^{\ast}$ and $L_{\text{max}}=10^{2}L_0^{\ast}$.

Similarly the Lorentz factor's dependence over angle follows:
\begin{equation}
\Gamma(\theta) = (\Gamma_{0}-1)y_{\Gamma}(\theta) + 1,
\end{equation}
with $\Gamma_{0}$ being the Lorentz factor of the jet at $\theta=0$.
Given these definitions, both $y_{L}$ and $y_{\Gamma}$ are defined to equal 1 at $\theta=0$.

The Lorentz factor determines the degree of relativistic beaming, which for the luminosity is governed by:
\begin{equation}\label{equ:relbeam}
\begin{aligned}
\mathcal{B}_L(\theta_{v}, \theta) =& \frac{1}{4\Gamma(\theta)^6(A^{2}-B^{2})^{2}}\\
& \left(5\left(\frac{A}{\sqrt{A^2 - B^2}}\right)^3 + 3\left(\frac{A}{\sqrt{A^2 - B^2}}\right)\right),
\end{aligned}
\end{equation} 
with $A=1-\beta\cos\theta\cos\theta_v$ and $B=-\beta\sin\theta\sin\theta_v$ where $\beta = \sqrt{1-\Gamma(\theta)^{-2}}$.

The apparent isotropic equivalent luminosity, for an observer at $\theta_v$ from the jet axis, can then be related to the intrinsic luminosity via the beaming function by combining Eqn.~\ref{equ:intrinsicE} and Eqn.~\ref{equ:relbeam}:
\begin{equation}
L_{\text{iso}}(\theta_{v}) = L_{0}\int_{0}^{\theta_j}\frac{1}{2} \mathcal{B}(\theta_{v},\theta)y_{L}(\theta)\sin\theta d\theta,
\label{equ:Eisoy}
\end{equation}
where $\theta_j$ is the maximum angle for which the beaming of gamma-rays occurs.
A conservative maximum outer jet angle for the emission of gamma-rays is approximated by considering scattering by electrons accompanying baryons within the jet.
The condition is given by \citep{matsumoto2019, lamb2022},
\begin{equation}\label{equ:thetaj}
\Gamma(\theta_j) \simeq 19.1\left(\frac{
L(\theta_j)}{10^{51}\,\text{erg/s}}\right)^{1/6},
\end{equation}
Beyond this limit, $\theta>\theta_j$, the jet becomes opaque to gamma-rays.

\subsection{Jet structures}\label{sec:JS}

The implications of structuring within compact stellar merger jets for the \ac{EM} counterparts from \ac{GW} detected systems has been highlighted in the literature \citep{lamb2017electromagnetic, lazzati2017off, kathirgamaraju2018, beniamini2020afterglow}; here we choose a sample of fiducial jet structure models that are representative of the literature diversity.

The top-hat jet (TH) is the simplest structure, where the beam is uniform until the jet opening angle $\theta_j$ where the jet sharply cuts off:
\begin{equation}
\begin{aligned}
y_{L}(\theta) =& \begin{cases} 1 & \text{ if } 0 \le \theta \le \theta_{j}, \\ 0 & \text{ else.} \end{cases}\\
y_{\Gamma}(\theta) =&\begin{cases} 1 & \text{ if } 0 \le \theta \le \theta_{j}, \\ 0 & \text{ else.} \end{cases}
\end{aligned}
\end{equation}
We note that the condition expressed in Eqn.~\ref{equ:thetaj} is not enforced for this case, as $\Gamma_0$ is above the Eqn.~\ref{equ:thetaj} limit at all points within the jet.

Wide-angle structure can be introduced with a Gaussian jet (GJ) structure, described by a single width parameter $\theta_{\sigma}$ \citep[e.g.][]{rossi2002afterglow, rossi2004polarization, zhang2002gamma, kumar2003evolution}:
\begin{equation}\label{equ:GJstruct}
y_{L}(\theta) = e^{-\frac{1}{2}\left(\frac{\theta}{\theta_{\sigma}}\right)^{2}},\quad
y_{\Gamma}(\theta) = e^{-\frac{1}{2}\left(\frac{\theta}{\theta_{\sigma}}\right)^{2}}.
\end{equation}
An alternative to the Gaussian profile has the wide-angle emission expressed as a three parameter power-law jet (PL) structure \citep[e.g.][]{kumar2003evolution, zhang2004quasi, rossi2004polarization}, where the jet can be described by some uniform core out to width $\theta_c$, and then the intrinsic luminosity structure falls off at wide angles according to power $s$ and the Lorentz factor with $a$:
\begin{equation}\label{equ:plaw}
\begin{aligned}
y_{L}(\theta) = & \begin{cases} 1 & \text{ if } 0 \le \theta \le \theta_{c}, \\ \left(\frac{\theta}{\theta_c}\right)^{-s} & \text{ else.}\end{cases}\\
y_{\Gamma}(\theta) = & \begin{cases} 1 & \text{ if } 0 \le \theta \le \theta_{c}, \\ \left(\frac{\theta}{\theta_c}\right)^{-a} & \text{ else.} \end{cases}
\end{aligned}
\end{equation}
Finally, let us consider a two component, or double Gaussian jet (DG), with emission from both an inner core described by a Gaussian structure of width $\theta_\text{in}$ and an outer cocoon described by width $\theta_\text{out}$ \citep{salafia2020}:
\begin{equation}
\begin{aligned}
y_{L}(\theta) = (1-\mathcal{C})e^{-\frac{1}{2}\left(\frac{\theta}{\theta_{\text{in}}}\right)^{2}} + \mathcal{C}e^{-\frac{1}{2}\left(\frac{\theta}{\theta_\text{out}}\right)^{2}},\\
y_{\Gamma}(\theta) = \frac{(1-\mathcal{C})e^{-\frac{1}{2}\left(\frac{\theta}{\theta_{\text{in}}}\right)^{2}} + \mathcal{C}e^{-\frac{1}{2}\left(\frac{\theta}{\theta_{\text{out}}}\right)^{2}}}{(1-\frac{\mathcal{C}}{\mathcal{A}})e^{-\frac{1}{2}\left(\frac{\theta}{\theta_{\text{in}}}\right)^{2}} + \frac{\mathcal{C}}{\mathcal{A}}e^{-\frac{1}{2}\left(\frac{\theta}{\theta_{\text{out}}}\right)^{2}}}.
\end{aligned}
\end{equation}
The luminosity of the outer cocoon is equal to $\mathcal{C}L_{0}$ and the Lorentz factor $\mathcal{A}(\Gamma_{0}-1) + 1$.

We do not consider hollow-cone jet structure models in our study \citep[see e.g.,][]{nathanail2021, takahashi2021}; we expect that the combination of our intrinsic luminosity distribution (Eqn.~\ref{equ:pLsig}) and the beaming (Eqn.~\ref{equ:Eisoy}) will wash-out the effect of any hollow-cone structuring within the core.
For this study, the structure outside of the jet's core is the critical component.

\section{Bayesian framework}\label{sec:anal}

Constraints are placed on model parameters $\lambda$ of a model $M$ when given data $\mathcal{D}$ in Bayesian data analysis by determining the posterior distribution using \textit{Bayes theorem}:
\begin{equation}\label{equ:Bayes}
    p(\lambda|\mathcal{D},M)=\frac{\mathcal{L}(\mathcal{D}|\lambda)\pi(\lambda)}{p(\mathcal{D}|M)},
\end{equation}
where $\mathcal{L}$ is the likelihood, $\pi$ is the prior and the normalisation term $p(\mathcal{D}|M)$ is the evidence.
Consider comparing two models $M_{1}$ and $M_{2}$ when given data $\mathcal{D}$. 
In the context of Bayesian data analysis, the statistic used to compare two models is the \textit{posterior odds} defined:
\begin{equation}
\mathcal{O}_{12} = \frac{p(M_{1}|\mathcal{D})}{p(M_{2}|\mathcal{D})} = \frac{p(M_{1})}{p(M_{2})}\frac{p(\mathcal{D}|M_{1})}{p(\mathcal{D}|M_{2})}.
\end{equation}
Normally we are interested in cases where the \textit{a priori} probability of either model being correct is comparable, and therefore the posterior odds is dominated by the \textit{Bayes factor}:
\begin{equation}\label{equ:BF}
\mathcal{B}_{12} = \frac{p(\mathcal{D}|M_{1})}{p(\mathcal{D}|M_{2})},
\end{equation}
which quantifies the contribution to the posterior odds given by the data $\mathcal{D}$.
A value of $\ln\mathcal{B}_{12}>0$ favours $M_{1}$, while $\ln\mathcal{B}_{12}<0$ favours $M_{2}$.

The analysis is performed by applying the model described in Section~\ref{sec:BG} with an assumed jet structure from Section~\ref{sec:JS} given both \ac{GW} and \ac{sGRB} prompt emission data.

Table~\ref{tab:varname} lists the notation used in the following section.
The data can be split into that produced by a \ac{GW}-triggered event, denoted with the subscript `GW', and that produced from an \ac{EM} trigger, denoted with the subscript `EM'.

The data from the $N_{\text{GW}}$ \ac{GW}-triggered events consists of the \ac{GW} strain $x_{\text{GW}}$ and the flux of the counterpart $F_{\text{GW}}$.
The \ac{GW}-triggered events may not necessarily require a counterpart to be considered for the analysis. 
If the sky localisation of the source coincides with the sky coverage of gamma-ray burst detectors then we can assume that it was not detected due to its distance and orientation to us.
The current events that meet this criteria are both GW170817 with GRB 170817A as well as GW190425 and the non-detection of its counterpart, under the assumption that a \ac{sGRB} was produced, given the Fermi detector covered $50\%$ of the sky localisation and Konus–\textit{Wind} covered the entire sky~\citep{hosseinzadeh2019follow}.

The \ac{EM}-triggered events are simply the number of \ac{sGRB} detections that Swift made within a 10 year operational period $N_{\text{EM}}$.

\begin{table}[!ht]
	\begin{center}
		\begin{tabular}{c|c}
			\hline
			\hline
			Variable & Description \\
			\hline
			$x_{\text{GW}}$ & GW detector data\\
			$F_{\text{GW}}$ & sGRB detector data\\
    		$N_{\text{GW}}$ & Number of detected GWs \\
			$N_{\text{EM}}$ & Number of detected sGRBs \\
			$\Sigma$ & Luminosity function hyperparameters\\
			$R_{\text{BNS}}$ & BNS merger rate \\
			$\Theta$ & Jet structure parameters \\
			$\Phi$ & $\{\theta_{v}, d_{L}\}$ \\
			$L_{0}$ & Intrinsic on-axis luminosity \\
			\hline
	    \end{tabular}
    \end{center}
	\caption{Shorthand notation of the \ac{GW} and \ac{EM} data as well as sets of parameters of interest.\label{tab:varname}}
\end{table}

\begin{table}[!ht]
	\begin{center}
		\begin{tabular}{c|c}
			\hline
			\hline
			Data set & Data \\
			\hline
            $\mathcal{D}_{170817}$ & $\{x_{170817}, F_{170817}\}$ \\
    		$\mathcal{D}_{190425}$ & $\{x_{190425}, F_{190425}\}$ \\
			$\mathcal{D}_{R}$ & $\{N_{\text{EM}},N_{\text{GW}}\}$\\
			$\mathcal{D}_{170817+R}$ & $\{\mathcal{D}_{170817},\mathcal{D}_{R}\}$ \\
			$\mathcal{D}_{\text{all}}$ & $\{\mathcal{D}_{170817},\mathcal{D}_{190425},\mathcal{D}_R\}$ \\
			\hline
	\end{tabular}
    \end{center}
	\caption{Summary of the data used in the analysis.}
	\label{tab:datalabel}
\end{table}

\begin{figure}[!ht]
	\begin{center}
		\scalebox{1.0}{
			\begin{tikzpicture}
			\node[obs] (xem) {$F_{\text{GW}}$};
			\node[obs, right=of xem] (xgw) {$x_{\text{GW}}$};						
			\node[latent, above=of xgw] (Phi) {$\Phi$};
			\node[latent, above left=of xem] (Theta) {$\Theta$};
			\node[right=of xgw] (dgw) {};
			\node[left=of Theta] (dem) {};
			\node[obs, above=of dgw, yshift=1.3cm] (Ngw) {$N_{\text{GW}}$};
			\node[obs, above=of dem, yshift=0.1cm] (Nem) {$N_{\text{EM}}$};
			\node[latent, above right=of Nem, xshift=2.0cm] (R0) {$R_{\text{BNS}}$};
			\node[latent, above =of xem] (E0) {$L_{0}$};
			\node[latent, above =of E0, yshift=-0.3cm] (lambda) {$\Sigma$};
			\edge {Theta,Phi,E0} {xem};
			\edge {Phi} {xgw};
			\edge {lambda} {E0};
			\edge {R0,xgw} {Ngw};
			\edge {R0,Theta,lambda} {Nem};
			\plate {plate} {(xem)(xgw)(Phi)} {$\forall\ i \in N_{\text{GW}} $} ;
			\end{tikzpicture}
		}
	\end{center}
	\caption{High-level Bayesian network of the model described in Section~\ref{sec:BG}. The variable names are defined in Table~\ref{tab:varname}.}
	\label{fig:BN}
\end{figure}
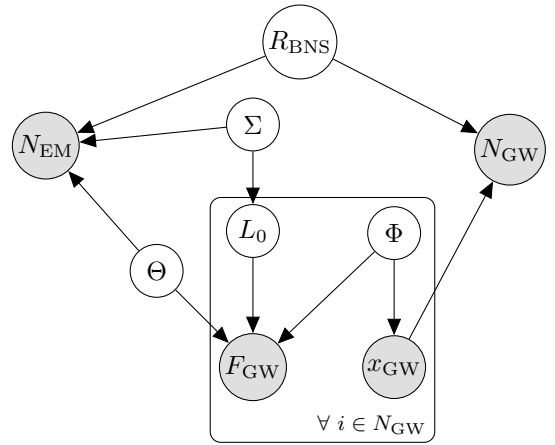

The likelihood can be decoupled into two terms, one of which considers \ac{GW}-triggered events and the other \ac{EM}-triggered:
\begin{equation}
\mathcal{L} = \mathcal{L}_{\text{EM}}\mathcal{L}_{\text{GW}}.
\end{equation}
The likelihood of the \ac{EM}-triggered events is a Poisson distribution with a mean given in Eqn.~\ref{equ:EMN}:
\begin{equation}
\mathcal{L}_{\text{EM}} = \frac{\hat{N}_{\text{EM}}(\Theta,R_{\text{BNS}},\Sigma)^{N_{\text{EM}}}e^{-\hat{N}_{\text{EM}}(\Theta,R_{\text{BNS}},\Sigma)}}{N_{\text{EM}}!}.
\end{equation}
The mean is evaluated over a regular grid of shape $(z, (2\theta/\pi)^{1/3}, \log_{10} L_{0})=(\times50,\times100,\times1000)$.
The angular grid points were chosen to be distributed over a power-law so as to populate low $\theta$ areas of the parameter space with grid points, while also maintaining a relatively high density of points at wider angles where emission from some jet structures is still significant.

The \ac{GW}-triggered events likelihood is the product of each of the $N_{\text{GW}}$ events:
\begin{equation}
\begin{aligned}
\mathcal{L}_{\text{GW}} \propto \prod_{i=1}^{N_{\text{GW}}}\sum_{j=1}^{S} p(F_{i,\text{GW}}|\Theta,\Phi_{i,j},L_{j,0}),
\end{aligned}
\end{equation}
where $S$ samples are taken of $\Phi_{i,j}$ and $L_{j,0}$ from $p(\Phi_{i},L_{0}|x_{i,\text{GW}},\Sigma)$.
The parameters can be sampled from separate distributions $p(\Phi_{i}|x_{i,\text{GW}})$ and $p(L_{0}|\Sigma)$ respectively, where $p(\Phi_{i}|x_{i,\text{GW}})$ are samples from the posteriors produced from \ac{GW} parameter estimation for each event.
The likelihood of the prompt emission of the \ac{GW}-triggered events is assumed to be a Gaussian distribution of width $\sigma_{F}$ about a mean described in Eqn.~\ref{equ:flux}:
\begin{equation}\label{equ:EML}
p(F_{\text{GW}}|\Theta,\Phi,L_{0}) = \frac{1}{\sqrt{2\pi\sigma_{F}^{2}}}\exp\left(-\frac{(F_{\text{GW}}-F)^{2}}{2\sigma_{F}^{2}}\right).
\end{equation}
The priors for the model are specified in Table~\ref{tab:priors} for each of the gamma-ray burst rate, luminosity function and jet structure parameters.
A normal distribution is denoted $\mathcal{N}(\mu,\sigma)$ with a mean of $\mu$ and standard deviation of $\sigma$, a uniform distribution as $\mathcal{U}(A,B)$ with lower bound of $A$ and upper bound of $B$, a Gamma distribution and inverse Gamma distribution as $\Gamma(\alpha,\beta)$ and $\Gamma^{-1}(\alpha,\beta)$ with a shape of $\alpha$ and a scale of $\beta$.
We assume that every \ac{BNS} merger results in a gamma-ray burst so that $\epsilon_{\text{BNS}}=1$.

\begin{table}[!ht]
	\centering
		\begin{tabular}{c|c|c}
			\hline
			\hline
			Model & Parameter & Prior \\
			\hline
			 & $\log_{10}R_{\text{BNS}}'$ & $\mathcal{N}(-6.6,0.77)$ \\
             \hline
             & $\log_{10}L_{0}^{*}$$'$ & $\mathcal{N}(51.6,1)$ \\
             & $\gamma$ & $\mathcal{U}(0,1)$ \\
             \hline
             & $\log_{10}\Gamma_{0}$ & $\textit{Gamma}^{-1}(2,2.5\times10^{-3})$ \\
             \hline
            TH & $\theta_{j}$ & $\mathcal{U}(0,\pi/2)$ \\
             \hline
            GJ & $\theta_{\sigma}$ & $\mathcal{U}(0,\pi/2)$ \\
             \hline
            \multirow{3}{*}{PL} & $\theta_{c}$ & $\mathcal{U}(0,\pi/2)$ \\
             & $s$ & $\textit{Gamma}(2,4)$ \\
             & $a$ & $\textit{Gamma}(2,1)$ \\
             \hline
            \multirow{4}{*}{DG} & $\theta_{\text{in}}$ & $\mathcal{U}(0,\pi/2)$ \\
             & $\theta_{\text{out}}$ & $\mathcal{U}(0,\theta_{\text{out}})$ \\
             & $\log_{10}\mathcal{C}$ & $\mathcal{U}(-6,0)$ \\
             & $\log_{10}\mathcal{A}$ & $\mathcal{U}(-6,0)$ \\
			\hline
	\end{tabular}
	\caption{Assumed prior distributions for each parameter. The analysis assumes one jet structure out of the top-hat (TH), Gaussian (GJ), power-law (PL) and double Gaussian (DG) models. Some parameters are made unit-less so that $R_{\text{BNS}}'=R_{\text{BNS}}/\text{Mpc}^{-3}\text{\,yr}^{-1}$ and $L_{0}^{*}$$'$$=L_{0}^{*}/\text{erg\,s}^{-1}$.}
	\label{tab:priors}
\end{table}

The T90 integrated flux of GRB\,170817A in the Fermi detector's $50-300$\,keV band is set at $F_{170817}=1.4{\times}10^{-7}$\,erg\,s$^{-1}$\,cm$^{-2}$ with an uncertainty of $\sigma_{170817}=3.64{\times}10^{-8}$\,erg\,s$^{-1}$\,cm$^{-2}$~\citep{goldstein2017ordinary}.
For the unobserved counterpart of GW190425, it is assumed that the T90 integrated flux takes a value of zero with an uncertainty of $\sigma_{190425}=10^{-8}$\,erg\,s$^{-1}$\,cm$^{-2}$ as a conservative upper bound to the Fermi detector's detection threshold~\citep{tan2020jet}.
The distance and viewing angle posteriors of GW170817 and GW190425 are each represented by $500$ samples taken from the their respective parameter estimation data releases.
In this work we consider an observing period of approximately $9.8$ years by the Swift detector in which it observed $N_{\text{EM}}=107$ \ac{sGRB}s as recorded by \citet{lien2016third}, given a sky coverage of $\Delta\Omega = 0.1$.
The log prior on the rate of \ac{BNS} mergers of $\mathcal{N}(-6.6,0.77)$ is roughly chosen to reflect the constraints imposed to the rates by GWTC-2~\citep{abbott2020population}.
The log prior on $L_0^{\ast}$ is centred around the fitted value taken from \citet{mogushi2019jet} with a standard deviation set to span one order of magnitude.
This is chosen to reflect some prior information in the allowed luminosity from prior observations, but with a width to allow for flexibility into higher or lower luminosity regimes.

Posteriors and Bayes factors are calculated from Eqn.~\ref{equ:Bayes} and Eqn.~\ref{equ:BF} by assigning $\mathcal{D}$ and $\lambda$ as the variables in Table~\ref{tab:varname}.
We collect the data into three sets: one only given the number of Swift  detections $N_{\text{EM}}$ and \ac{GW} detections $N_{\text{GW}}$ called $\mathcal{D}_{R}$, another with the combined GW170817 \ac{GW} $x_{170817}$ and \ac{EM} data $F_{170817}$ called $\mathcal{D}_{170817}$, and the other with GW190425 \ac{GW} data $x_{190425}$ and the flux from the non-detection $F_{190425}$ called $\mathcal{D}_{190425}$. 
The analysis is performed on five combinations of these data sets: $\mathcal{D}_{170817}=\{x_{170817},F_{170817}\}$, $\mathcal{D}_{190425}=\{x_{190425},F_{190425}\}$, $\mathcal{D}_{R}=\{N_{\text{EM}},N_{\text{GW}}\}$, $\mathcal{D}_{170817+R}=\{\mathcal{D}_{170817},\mathcal{D}_{R}\}$, and $\mathcal{D}_{\text{all}}=\{\mathcal{D}_{170817},\mathcal{D}_{190425},\mathcal{D}_{R}\}$.
We allow that $\lambda=\{\Sigma,\mathcal{R}_0,\Theta_{M}\}$, where $\Theta_M$ are the jet structure model parameters dependent on jet structure model $M$.
The three analyses are repeated for each of the jet structure models: $M=$\,TH, GJ, PL and DG.

The posterior samples and evidence for each case are calculated via the nested sampling algorithm \textsc{Nessai}, that utilises machine learning techniques to drastically reduce the number of evaluations of the expensive likelihood function~\citep{williams2021nested}.

\section{Results}\label{sec:results}

\begin{figure*}[!ht]
\gridline{\fig{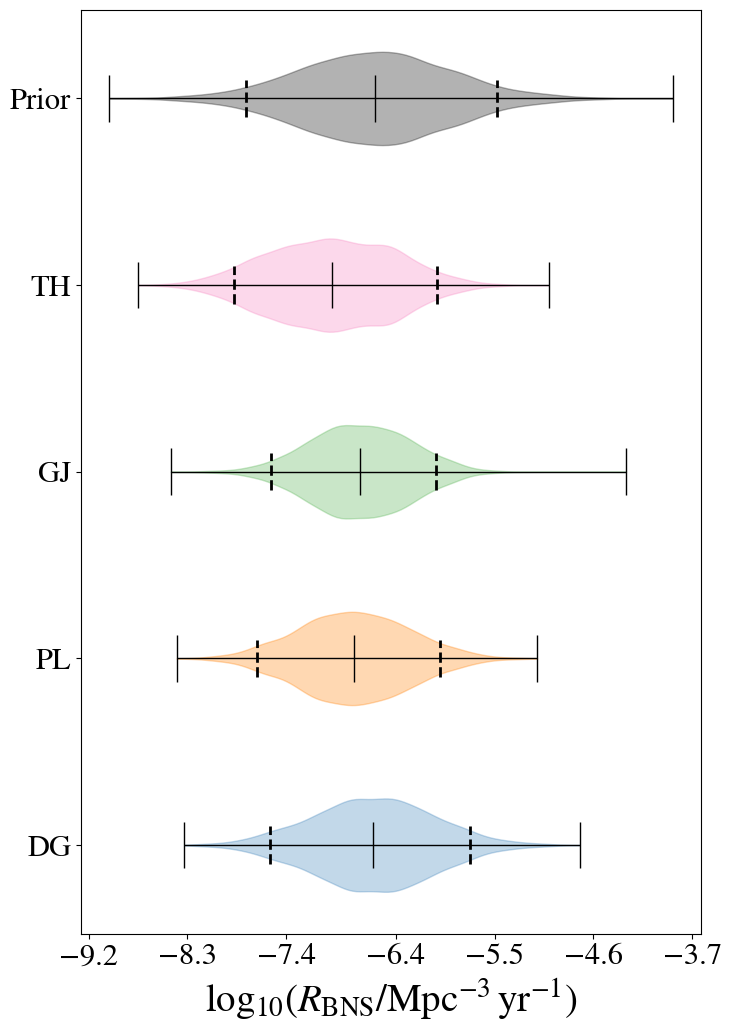}{0.45\textwidth}{(a)\label{fig:BNSrate}}
          \fig{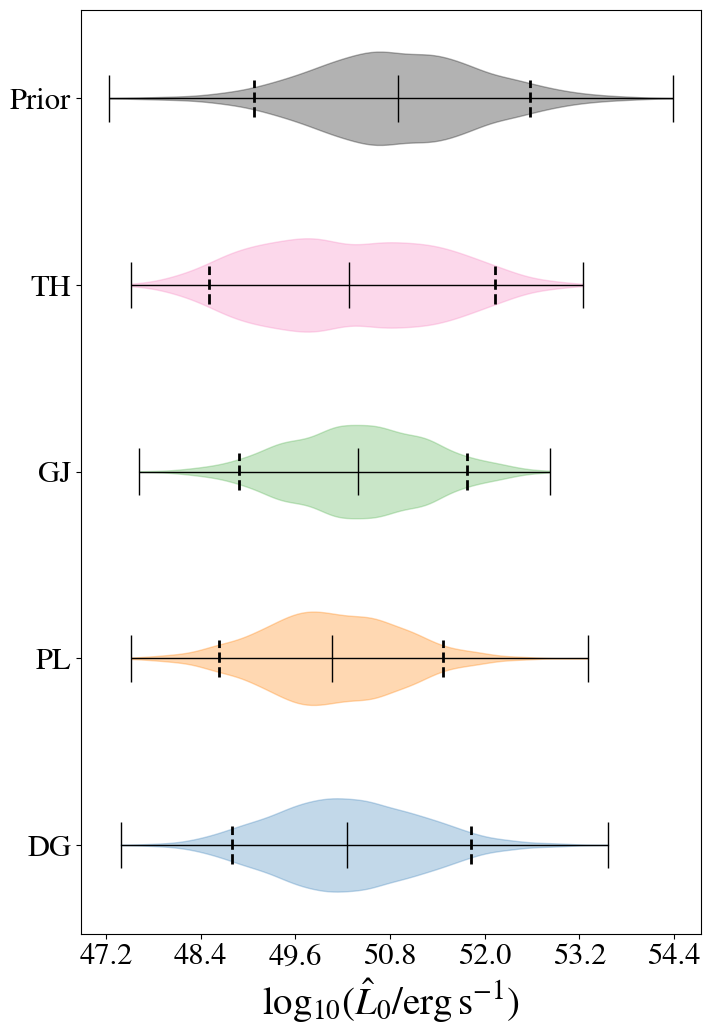}{0.45\textwidth}{(b)\label{fig:L0}}}
	\caption{Posterior distributions of $(a)$ the rate of \ac{BNS} mergers and $(b)$ the mean intrinsic luminosity $\hat{L}_{0}$ assuming the top-hat jet (TH, pink), Gaussian jet (GJ, green), power-law jet (PL, orange), and double Gaussian jet (DG, blue) models when given $\mathcal{D}_{\text{all}}$. The solid vertical lines represent the minimum and maximum of each distribution while the thickness of the fill in between represents the probability density. The dashed vertical lines represent $90\%$ credible intervals. The median is shown by the middle solid line. The top distribution represents the prior distribution taken from Table~\ref{tab:priors} for $\log_{10}R_{\text{BNS}}$, $L_0$ and $\gamma$ respectively. Similar posterior distributions are recovered for models with wide-angled structuring in comparison to the top-hat model case.}
	\label{fig:rateL0}
\end{figure*}

\begin{table*}[!ht]
	\resizebox{1.9\columnwidth}{!}{%
		\begin{tabular}{c|cccc}
			\hline
			\hline
			$\ln\mathcal{B}_{\text{row,col}}$ & Top-hat & Gaussian & Power-law & Double Gaussian \\
			\hline
            Top-hat & $0$ & $\sfrac{-0.54}{-0.91}$ & $\sfrac{-0.46}{-0.45}$ & $\sfrac{-0.55}{-0.64}$ \\
            Gaussian & $\sfrac{0.54}{0.91}$ & $0$ & $\sfrac{0.08}{0.46}$ & $\sfrac{-0.01}{0.27}$ \\
            Power-law & $\sfrac{0.46}{0.45}$ & $\sfrac{-0.08}{-0.46}$ & $0$ & $\sfrac{-0.09}{-0.19}$ \\
            Double Gaussian & $\sfrac{0.55}{0.64}$ & $\sfrac{0.01}{-0.27}$ & $\sfrac{0.09}{0.19}$ & $0$ \\
			\hline
	\end{tabular}}
	\caption{The log Bayes factor $\ln\mathcal{B}$ between each model when given $\mathcal{D}_{\text{all}}$ $\sfrac{\text{w/o}}{\text{with}}$ the fitted luminosity function. A positive value is evidence towards the model of the row while a negative is evidence that favours the column model. Slight evidence is provided for models with wide-angled structuring over the top-hat model while little evidence distinguishes between the power-law, Gaussian and double Gaussian models without the fitted luminosity function. When the fitted luminosity function is included, slight evidence is further provided in favour of the Gaussian jet over the double Gaussian and power-law jet structures while the top-hat remains least favoured. All values can be assumed to have uncertainties of $\pm 0.05$.}
	\label{tab:lnBF}
\end{table*}

The analysis that is described in the previous section is applied to all three sets of data.
The full corner plots for each jet structure model are shown in the appendix, where Figure~\ref{fig:TH} shows the results for the top-hat jet, Figure~\ref{fig:GJ} the Gaussian jet, Figure~\ref{fig:PL} the power-law jet and Figure~\ref{fig:DG} the double Gaussian jet structure model.
The posteriors for each of the data sets is overlaid upon one another where $\mathcal{D}_{170817}$ is shown in red, $\mathcal{D}_{190425}$ in orange, $\mathcal{D}_{R}$ in violet, $\mathcal{D}_{170817+R}$ in blue and $\mathcal{D}_{\text{all}}$ in black.
The log evidence $\ln p(\mathcal{D}|M)$ that corresponds to each posterior is shown in Table~\ref{tab:lnZ} for each of the five data sets over the four jet structure models.
The log Bayes factors between the different models given the same data set can simply be calculated by taking the difference between entries of the same row.

Much of the discussion in this section concerns the posterior constraints when all of the data is considered $\mathcal{D}=\mathcal{D}_{\text{all}}$, however the outcomes given the other subsets are considered to explain these results and provide further insight.

The log Bayes factors between the jet structure models are given as the left-hand entries of Table~\ref{tab:lnBF} when given the $\mathcal{D}_{\text{all}}$ data.
A positive value indicates that the data supports the model of the row while a negative value supports the column model.
Evidently the top-hat model is less favourable than the models with wide-angled jet structuring, with Bayes factors of $-0.54$, $-0.46$ and $-0.55$ between it and the Gaussian, power-law and Double Gaussian jet structures respectively.
The log Bayes factors between the power-law, Gaussian and the double Gaussian is slight, with only insignificant evidence in favour of the Gaussian and double Gaussian model of log Bayes factors of less than $0.1$, and negligibly small log Bayes factors between the two.

The constraints on the rate of \ac{BNS} mergers when given $\mathcal{D}_{\text{all}}$ are shown in Figure~\ref{fig:BNSrate} for the four jet structure models.
These constraints take the form of posterior distributions that are represented in the violin plots, where the outermost solid vertical lines indicate the minimum and maximum sample value while the fill in between represents the probability density.
The $90\%$ narrowest credible intervals are shown by the vertical dashed lines which enclose the median indicated by the middle solid line.
The posterior distributions are compared to samples from the prior distribution at the top of the figure.
For all cases, the posterior places tighter constraints on the merger rate than the prior distribution.
The cases with wide-angled structuring (Gaussian, power-law and double Gaussian models) produce similar posterior distributions to one another, centred around a value of ${\sim} 10^{-7}\,$Mpc$^{-3}\,$yr$^{-1}$ consistent with the mean of the prior.
The Gaussian jet structure produces the narrowest constraints with a $90\%$ credible interval of $\log_{10}R_{\text{BNS}}/\text{Mpc}^{-3}\,\text{yr}^{-1}=-6.7^{+0.7}_{-0.8}$, compared to the power-law and double Gaussian models of $-6.8^{+0.7}_{-0.9}$ and $-6.6^{+0.9}_{-0.9}$ respectively.
The top-hat model favours lower rates of \ac{BNS} mergers, and even pushes the lower bound on the $90\%$ credible interval to lower values of that of the prior, constraining it between $\log_{10}R_{\text{BNS}}/\text{Mpc}^{-3}\,\text{yr}^{-1}=-7^{+0.9}_{-1.0}$.

The median intrinsic luminosity posteriors determined for each model when given $\mathcal{D}_{\text{all}}$ is shown in Figure~\ref{fig:L0}.
The mean intrinsic luminosity $\hat{L}_{0}$ is determined by drawing $L_0$ and $\gamma$ from the respective posterior distribution and then drawing $1000$ samples from the corresponding Schechter function of Eqn.~\ref{equ:pLsig} before finding the ensemble median.
This process is then repeated for $2000$ median intrinsic luminosity samples.
These posteriors take a form similar to the rates posteriors in Figure~\ref{fig:BNSrate} as violin plots where the shaded probability density is contained within the outermost maximum and minimum values indicated by the solid vertical lines, while the median is marked by the middle solid line.
The median is enclosed by the narrowest $90\%$ credible intervals displayed as dashed vertical lines.
A distribution of prior samples of $\hat{L}_0$ is also plotted at the top of the figure, which is determined by sampling from the individual $\log_{10}L_{0}^{*}$ and $\gamma$ priors defined in Table~\ref{tab:priors}.
The top-hat jet structure resembles the prior in width, but shifts to favour lower luminosity and exhibits some bimodality as the probability density pinches at the median.
This is due to the bimodality of the $L_0$ posterior distribution in Figure~\ref{fig:TH} given $\mathcal{D}_{170817+R}$, which shall be discussed later in Section~\ref{sec:disc}.
The models with wide-angled structure tend towards lower mean intrinsic luminosity values, with the Gaussian model constrained to $\log_{10}\hat{L}_0/\text{erg}\,\text{s}^{-1} = 50.4^{+1.6}_{-1.3}$, power-law model $50.1^{+1.3}_{-1.6}$ and double Gaussian model of $50.2^{+1.5}_{-1.5}$.
The top-hat model is constrained to $\log_{10}\hat{L}_0/\text{erg}\,\text{s}^{-1} = 50.2^{+1.7}_{-1.7}$ which we can compare to the prior of $51^{+1.8}_{-1.8}$.

\begin{table}[!ht]
	\centering
		\begin{tabular}{c|c|c|c}
			\hline
			\hline
			Model & Parameter & \multicolumn{2}{|c}{Constraints} \\
			 &  & w/o fitted LF & Fitted LF \\
			\hline
            TH & $\theta_{j}$ & $14.9^{+46.0}_{-14.3}\phantom{ }^{\circ}$ & $9.2^{+17.1}_{-7.9}\phantom{ }^{\circ}$ \\
             \hline
            GJ & $\theta_{\sigma}$ & $5.9^{+28.2}_{-5.4}\phantom{ }^{\circ}$ & $4.2^{+5.2}_{-3.2}\phantom{ }^{\circ}$ \\
             \hline
            \multirow{3}{*}{PL} & $\theta_{c}$ & $10.6^{+32.9}_{-10.0}\phantom{ }^{\circ}$ & $6.0^{+9.2}_{-4.9}\phantom{ }^{\circ}$ \\
             \cline{2-4}
             & $s$ & $6.7^{+8.1}_{-5.2}$ & $6.4^{+7.9}_{-4.3}$ \\
             \cline{2-4}
             & $a$ & $1.5^{+2.2}_{-1.4}$ & $1.5^{+2.1}_{-1.4}$ \\
             \hline
            \multirow{4}{*}{DG} & $\theta_{\text{in}}$ & $6.3^{+24.1}_{-6.2}\phantom{ }^{\circ}$ & $3.6^{+6.2}_{-2.8}\phantom{ }^{\circ}$ \\
             \cline{2-4}
             & $\theta_{\text{out}}$ & $49.3^{+40.7}_{-38.0}\phantom{ }^{\circ}$ & $46.2^{+43.1}_{-36.3}\phantom{ }^{\circ}$ \\
             \cline{2-4}
             & $\log_{10}\mathcal{C}$ & $-3.8^{+2.9}_{-2.2}$ & $-4.1^{+1.9}_{-1.9}$ \\
             \cline{2-4}
             & $\log_{10}\mathcal{A}$ & $-3.0^{+2.8}_{-2.9}$ & $-3.1^{+2.9}_{-2.7}$\\
			\hline
	\end{tabular}
	\caption{Constraints placed on each of the variables for each model given data set $\mathcal{D}_{\text{all}}$ with and without the fitted luminosity function (LF). The median of each posterior distribution is quoted along with upper and lower bounds placed by $90\%$ credible intervals.}
	\label{tab:JSpost}
\end{table}

The constraints from the posteriors on the jet structure parameters given each jet structure model are shown in Table~\ref{tab:JSpost}.
The median is quoted along with the upper and lower bounds placed by the $90\%$ narrowest credible intervals.

\subsection{Fitted luminosity function}

The analysis is repeated but instead of assuming a prior distribution on the luminosity scale and shape, a luminosity function fitted from the observed isotropic equivalent luminosity of \ac{sGRB}s with associated redshifts.
The values of $L_0$ and $\gamma$ are taken from the mean fitted Schechter function in \citet{mogushi2019jet} of $\log_{10}L_0/\text{erg}\,\text{s}^{-1}=51.6$ and $\gamma=0.55$, fitted to the isotropic equivalent luminosity of $35$ \ac{sGRB}s.

The log evidence between each of the jet structure models and the different data sets are shown in Table~\ref{tab:lnZFLF}, while the respective Bayes factors when given $\mathcal{D}_{\text{all}}$ between each of the jet structure models are shown on the right-hand entries of Table~\ref{tab:lnBF}.

\begin{figure}[!ht]
	\centering
	\includegraphics[width=0.45\textwidth]{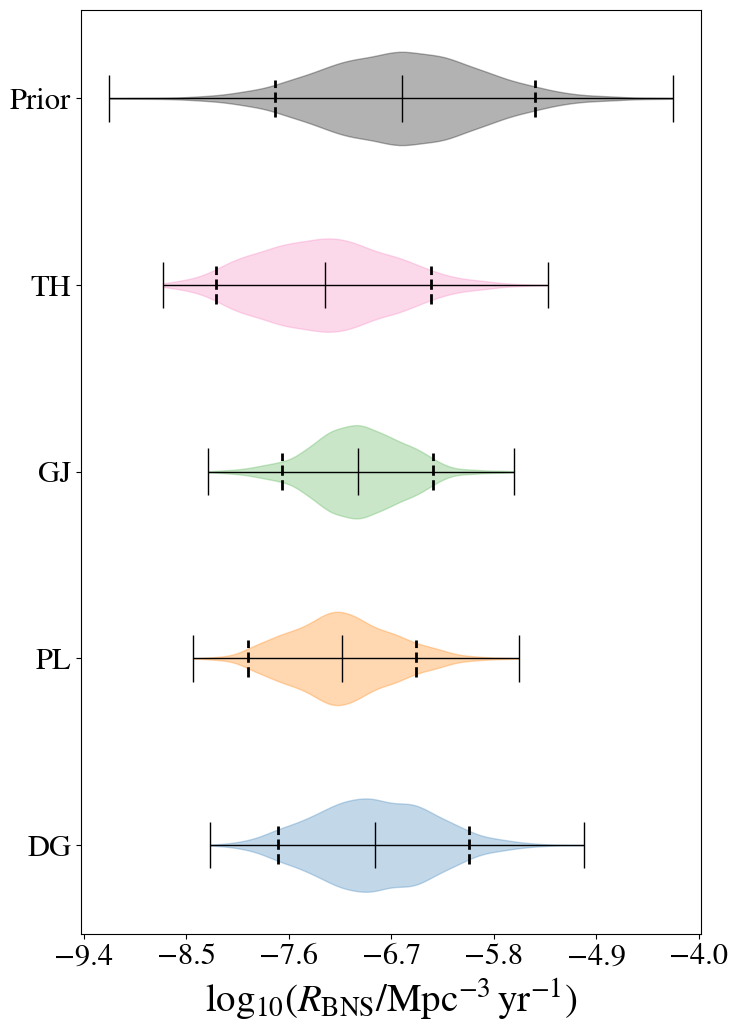}
	\caption{Posterior distributions of the rate of \ac{BNS} mergers assuming the top-hat jet (TH, pink), Gaussian jet (GJ, green), power-law jet (PL, orange) and double Gaussian jet (DG, blue) models when given $\mathcal{D}_{\text{all}}$ and the fitted luminosity function. The solid vertical lines represent the minimum and maximum of each distribution while the thickness of the fill in between represents the probability density. The dashed vertical lines represent $90\%$ credible intervals. The median is shown by the middle solid line. The posterior distributions are compared to samples from the prior of $\mathcal{N}(-6.6,0.77)$ at the top. All models recover similar constraints that narrow about the mean value assigned to the prior.} 
	\label{fig:BNSrateFLF}
\end{figure}

The posteriors on the local rate of \ac{BNS} merger when given the fitted luminosity function and $\mathcal{D}_{\text{all}}$ are shown in Figure~\ref{fig:BNSrateFLF} in the same format as Figure~\ref{fig:BNSrate}, where the widths of the violin plots indicate the probability density, the maximum and minimum sample is indicated by the extreme solid vertical lines, and median with the middle solid vertical line.
The narrowest $90\%$ credible intervals are indicated by the dashed vertical lines and are 
$\log_{10}R_{\text{BNS}}/\text{Mpc}^{-3}\,\text{yr}^{-1}=-7.0^{+0.7}_{-0.6}$, $-7.2^{+0.7}_{-0.8}$, $-6.8^{+0.8}_{-0.9}$ for the Gaussian, power-law and double Gaussian models respectively.
For the top-hat jet model, the rate is constrained to $\log_{10}R_{\text{BNS}}/\text{Mpc}^{-3}\,\text{yr}^{-1}=-7.3^{+1.0}_{-0.9}$.

The constraints on the jet structure models when given the fitted luminosity function and $\mathcal{D}_{\text{all}}$ are presented in Table~\ref{tab:JSpost} for each jet structure models, with the upper and lower bounds representing the narrowest $90\%$ credible intervals.

\section{Discussion}\label{sec:disc}

\subsection{Wide-angle jet structuring}

The log Bayes factors are greater than $0.45$ for all models with wide-angle jet structure when compared to the top-hat model.
This is due to the top-hat jets failing to resolve the number of observed \ac{sGRB}s with the flux of GRB\,170817A when assuming that the event had a typical event opening angle.
Given the assumed star formation rate and the constraints on the \ac{BNS} merger rate from gravitational-wave detections, to obtain a Swift \ac{sGRB} detection rate of $11$\,yr$^{-1}$, the jets are either predicted to have narrow opening angles and high luminosities or wide opening angles and low luminosities.
This constraint can be seen in the bottom left corner plot panel of Figure~\ref{fig:TH} in the violet posterior, where much of the probability density is concentrated in the low luminosity and wide opening angle area of the parameter space.
In contrast, the constraints made by GW170817 and GRB\,170817A favour a wide opening angle of $\theta_j\lesssim\theta_v$ and a high luminosity event, as seen in the bottom left-hand panel of Figure~\ref{fig:TH} in the red posterior.
This is as emission from an event from a top-hat jet structure when viewed at wide angles can only come from Doppler beaming, which falls off very sharply with increased viewing angles.
As $\theta_{v}>\theta_{j}$ is more probable than $\theta_{v}\gtrsim\theta_{j}$, then a high luminosity event is deemed more probable.
The two constraints produce posteriors that share very little overlap in the parameter space, leading to the top-hat model providing a smaller evidence than the other models.
This contradiction also manifests in the bimodality of the mean luminosity posterior for the top-hat jet model, as seen in Figure~\ref{fig:L0}, as the lower luminosity high density region corresponds to the constraint produced from the observed number of gamma-ray bursts, while the higher luminosity high density region corresponds to the constraints made by GW170817 and GRB\,170817A.

Jet structure models with wide-angle structuring are only favoured when the data from GW170817/GRB\,170817A of $\mathcal{D}_{170817}$ is combined with the event rate information from $\mathcal{D}_{R}$.
When these data sets are considered individually the evidence for a top-hat jet structure is comparable or higher than the other models in most cases, as seen in the Table~\ref{tab:lnZ}. 
The inclusion of the GW190425 event provides evidence against jets with wide-angle structuring.
This can be seen by comparing the difference in log evidence between the top-hat jet and the other models given $\mathcal{D}_{170817+R}$, and the difference when given $\mathcal{D}_{\text{all}}$, where there is relatively less evidence between the models when $\mathcal{D}_{190425}$ is included.
As it is assumed that GW190425 produced a counterpart that went undetected due to its distance and viewing angle, the event places an upper-bound on the luminosity and jet width.
This upper bound on the jet structure limits the possible wide-angle emission, which makes wide-angle jet structuring unnecessary to explain the event.

\subsection{Cocoon emission}

The double Gaussian jet structure provides a stand-in for a jet structure with cocoon-like emission, where the outer Gaussian provides a secondary component for the emission contribution from an energetic cocoon.
Interestingly, comparing the log evidence given the double Gaussian jet model to the other models shows weak evidence for the double Gaussian jet structure when considering $\mathcal{D}_{170817}$, $\mathcal{D}_{R}$, $\mathcal{D}_{170817+R}$ for all cases (with the exception of $\ln p(\mathcal{D}_{170817+R}|\text{GJ})$ given the fitted Schechter luminosity function), suggesting it is the favourable model when considering both GW170817/GRB\,170817A and the observed rate data.
As discussed in the previous section, GW190425 places an upper-bound on the wide-angled emission and provides support for the top-hat and power-law jet structure with sharper cut-offs. 
Given that the suitability of GW190425 in the analysis is not as clear-cut as an event like GW170817 due to the uncertainty of the \ac{EM} coverage of the event, this result should not be disregarded.
While this may not provide convincing evidence for the observation of cocoon emission, it suggests that with the inclusion of future events, the necessity for the cocoon-like component can be better assessed.

\subsection{Rate of binary neutron star mergers}

The narrowest $90\%$ credible intervals of the rate of \ac{BNS} mergers are constrained within $1-1300$\,Gpc$^{-3}$\,yr$^{-1}$ independent of the jet structure model considered, improving upon the constraints imposed by GWTC-3~\citep{abbott2023population}.
The rate is further constrained to the interval of $2-680$\,Gpc$^{-3}$\,yr$^{-1}$ when the fitted Schechter luminosity function is assumed.
Future \ac{BNS} detections will provide tighter constraints on their merger rate.
These constraints will allow for a tighter prior to be placed on the rate of mergers, allowing the possible jet structures to be distinguished.

\subsection{Luminosity function}

Two different cases are explored in the analysis: one where the luminosity function is fitted in advance of the analysis, and the other where priors are placed on the luminosity function parameters $L_0^{\ast}$ and $\gamma$.
When priors are placed on the luminosity function, the luminosity function generally favours low luminosities for all models assumed.
This is apparent in Figure~\ref{fig:L0} where the $\hat{L}_0$ posterior for all jet structure models shifts to low luminosity when compared to the prior distribution, and mean values of $\log_{10}(L_{0}^{\ast}/\text{erg}\,\text{s}^{-1})$ shift to $51{-}51.25$ in comparison to the value of $51.6$ taken from \citet{mogushi2019jet} and used as the mean of the prior.

The inclusion of the fitted luminosity function informs the analysis of the prompt emission of all \ac{sGRB}s that are used in the fit --- information that is excluded from the case where the luminosity priors are placed.
This allows for narrower constraints on the jet structure model parameters, as seen by comparing the left to right hand-side of the last column of Table~\ref{tab:JSpost}.
Similarly, this also leads to tighter constraints in the \ac{BNS} merger rate posteriors as seen by comparing Figure~\ref{fig:BNSrate} to Figure~\ref{fig:BNSrateFLF}.
Interestingly, fitting the luminosity function provides slight evidence for the Gaussian jet structure model over all other jet structures given all the data, as seen by the right-hand log Bayes factors shown in Table~\ref{tab:lnBF}.
However, fitting the luminosity function requires assumptions about the jet structure to be made, which will lead to biases in this analysis.
In \citet{mogushi2019jet} which the fitted luminosity function is taken from, the fit is produced by assuming that all $35$ \ac{sGRB} prompt emission observations with associated redshifts are seen on-axis. 
However, if some of the events used in fitting the luminosity function where observed at an angle, then the observed variability in their observed isotropic luminosity would be wrongly attributed to variability in the intrinsic luminosity.
Assuming a wider distribution to the intrinsic luminosity would favour wider jet structures.
To avoid this bias, a future analysis should adjust the likelihood to accommodate the flux data of all observed \ac{sGRB}s along with their associated redshifts while placing priors on the luminosity function parameters.
This would allow for the luminosity function to be fitted internally within the analysis without having to make the additional jet structure assumptions in a pre-processing step.

\subsection{GW190425}

The inclusion of GW190425 in the analysis provides an upper bound to the wide-angled jet structure emission, due to the absence of an \ac{EM} detection. 
The viewing angle posterior of the event exhibits a similar distribution as that of GW170817, while the distance to the event is notably larger at a distance of approximately $160$\,Mpc compared to GW170817's distance of $40$\,Mpc.
The event is close enough in proximity that, if observed on-axis and is of typical luminosity, would produce a flux tens or hundreds of times greater than GRB\,170817A.
However, there are assumptions about the event that are made by including it in this way.
Firstly, it implies that the event produced a \ac{sGRB}.
This assumption is made explicitly in the analysis when incorporating the observed rate of merger, where every \ac{BNS} merger is assumed to produce a \ac{sGRB} in Eqn.~\ref{equ:localsGRB}.
However as the prior on the local rate of \ac{BNS} mergers is relatively wide and covers multiple orders of magnitude, this assumption should not affect the analysis when considering the whole population as long as \ac{BNS} mergers do typically produce \ac{sGRB}s.
This assumption has a much greater impact when analysing individual events where wrongly asserting a particular event produced a \ac{sGRB} leads to false conclusions.
Secondly, it is assumed that the event would be observed given a wider jet structure or higher luminosity.
While the event was within the field of view of the Konus-\textit{Wind} satellite, the incomplete sky coverage of the event by the more sensitive detectors such as the \textit{Fermi}-GBM detector and \textit{Swift}-BAT bring the detectability of the event into question. 
Despite the validity of these assumptions, and that the event produces evidence against wide-angle jet structuring, it is found that GW190425 is still compatible with the jet structure models given the rest of the data.
This can be assessed by the comparison of $\ln p(\mathcal{D}_{\text{all}}|M)$ to $\ln p(\mathcal{D}_{190425}|M)+\ln p (\mathcal{D}_{170817+R}|M)$ for each of the jet structure models $M$.
For all models, the value of $\ln p(\mathcal{D}_{\text{all}}|M)>p(\mathcal{D}_{190425}|M)+\ln p (\mathcal{D}_{170817+R}|M)$, suggesting that the observation of GW190425 is informative to the analysis in all cases, and does not conflict with the constraints imposed to the model given by the detection of GW170817/GRB\,170817A and the rate of observed \ac{sGRB}s.
This result suggests that it is feasible for GW190425 to have had a typical \ac{sGRB} counterpart
with the same jet structure as GRB\,170817A that would have remained undetectable to our instrumentation even given full sky coverage.
This observation is consistent with the result obtained in \citet{saleem2020energetics} where it was concluded that such a structured jet is consistent with the observed flux of the \textit{INTEGRAL} detector given the detector's flux upper limit.

\section{Conclusion}\label{sec:conc}

We provide an extensive Bayesian analysis that constrains the jet structure, intrinsic luminosity function and rate of \ac{BNS} mergers as well as providing a comparison between competing jet structure models.
This is achieved by combining four data avenues: 1. the parameter inference posteriors from a \ac{GW} trigger, 2. the \ac{sGRB} flux when a counterpart is detected or the detector flux upper limit otherwise, 3. the observation rate of detected \ac{sGRB}s, 4. the merger rate informed from \ac{GW} observations.
We perform this analysis using the \ac{GW} triggers GW170817 and GW190425, GRB\,170817A, the non-detection of a GW190425 counterpart, the rate of \ac{sGRB} detections by the Swift detector within a $9.8$ year observation period and a merger rate consistent with the constraints imposed by GWTC-2~\citep{abbott2020population}.
This provides us with the following results:
\begin{itemize}
    \item[$\bullet$] The rate of \ac{BNS} mergers is constrained within $1-1300$\,Gpc$^{-3}$\,yr$^{-1}$, improving upon the results of GWTC-3.
    \item[$\bullet$] Wide-angled jet structures prove more compatible with the given model than top-hat jet in explaining the observed number of \ac{sGRB}s in the wake of the low observed isotropic luminosity of GRB\,170817A.
    \item[$\bullet$] Slight evidence is provided for a cocoon-like wide-angled jet structure when considering the observed rate of \ac{sGRB}s and GRB\,170817A. However, the evidence becomes awash across all wide-angled jet structures when GW190425 is included in the analysis.
    \item[$\bullet$] While providing evidence against wide-angled structuring, the hypothesis that GW190425 had a typical \ac{sGRB} counterpart with a GRB\,170817A-like jet structure and would remain undetectable to the Fermi detector given full-sky coverage is feasible given the model. 
\end{itemize}

The analysis was extended to consider a fitted intrinsic luminosity function to further incorporate the detected flux and estimated redshifts of past \ac{sGRB} detections. 
This provides the results:
\begin{itemize}
    \item[$\bullet$] The rate of \ac{BNS} mergers is further constrained to $2-680$\,Gpc$^{-3}$\,yr$^{-1}$.
    \item[$\bullet$] Slight evidence for the Gaussian jet structure is provided, unless GW190425 is excluded in which the cocoon-like double Gaussian jet structure is equally favoured. 
\end{itemize}
However, we note that the fitting of the luminosity function requires strong assumptions about the jet structure and therefore introduces a bias towards jet structures with wide central components.
Interestingly, this bias does not appear to manifest in the resulting Bayes factors where the top-hat jet loses favour over the wide-angled jet structures.
A future analysis will work to incorporate the flux measurements and redshift estimations of detected \ac{sGRB}s directly, and therefore avoid introducing this bias.
Future work would also include incorporating afterglow data into the analysis for events that coincide with a \ac{GW} detection~\citep{lin2021bayesian}.

\section{Acknowledgements}

We are grateful for computational resources provided by Cardiff University, and funded by an STFC (grant no. ST/I006285/1) supporting UK Involvement in the Operation of Advanced LIGO.
The authors thank Shiho Kobayashi for fruitful discussions.
F.H. was supported by Science and Technology Research Council (STFC) (grant no. ST/N504075/1).
J.V., I.S.H. and M.J.W. are supported by STFC (grant no. ST/V005634/1).
M.J.W. was also supported by STFC (grant no. 2285031).
G.P.L. is supported by a Royal Society Dorothy Hodgkin Fellowship (grant no. DHF-R1-221175 and DHFERE-221005). 
E.T.L is supported by National Science and Technology Council (NSTC) of Taiwan (grant no. 111-2112-M-007-020).

\bibliography{sample631}{}
\bibliographystyle{aasjournal}

\onecolumngrid

\newpage
\appendix
\renewcommand{\thefigure}{A\arabic{figure}}
\setcounter{figure}{0}
\renewcommand{\thetable}{A\arabic{table}}
\setcounter{table}{0}

\section{Evidences}

\begin{table*}[!ht]
	\resizebox{.8\columnwidth}{!}{%
		\begin{tabular}{c|cccc}
			\hline
			\hline
			Data $\mathcal{D}$ & \multicolumn{4}{c}{Model $M$} \\
			& Top hat & Gaussian & Power-law & Double Gaussian \\
			\hline
			$\mathcal{D}_{170817}$ & $ 11.87 \pm 0.02$ & $ 11.8 \pm 0.03$ & $ 11.79 \pm 0.03$ & $ 12.08 \pm 0.03$ \\
			$\mathcal{D}_{190425}$ & $ 17.29 \pm 0.01$ & $ 16.3 \pm 0.03$ & $ 16.41 \pm 0.02$ & $ 16.42 \pm 0.03$ \\
			$\mathcal{D}_{R}$ & $ -7.67 \pm 0.04$ & $ -7.8 \pm 0.04$ & $ -7.73 \pm 0.04$ & $ -7.53 \pm 0.04$ \\
			$\mathcal{D}_{170817+R}$ & $ 4.45 \pm 0.05$ & $ 5.23 \pm 0.05$ & $ 5.15 \pm 0.05$ & $ 5.37 \pm 0.05$ \\
			$\mathcal{D}_{\text{all}}$ & $ 22.12  \pm 0.05$ & $ 22.66 \pm 0.05$ & $ 22.58 \pm 0.05$ & $ 22.67 \pm 0.05$ \\
			\hline
	\end{tabular}}
	\caption{The log evidence $\ln p(\mathcal{D}|M)$ given the five different data sets when assuming each of the four models.}
	\label{tab:lnZ}
\end{table*}

\begin{table*}[!ht]
	\resizebox{.8\columnwidth}{!}{%
		\begin{tabular}{c|cccc}
			\hline
			\hline
			Data $\mathcal{D}$ & \multicolumn{4}{c}{Model $M$} \\
			& Top hat & Gaussian & Power-law & Double Gaussian \\
			\hline
			$\mathcal{D}_{170817}$ & $ 11.48 \pm 0.02$ & $ 11.6 \pm 0.03$ & $ 11.45 \pm 0.03$ & $ 11.62 \pm 0.03$ \\
			$\mathcal{D}_{190425}$ & $ 17.28 \pm 0.01$ & $ 16.23 \pm 0.03$ & $ 16.29 \pm 0.03$ & $ 16.21 \pm 0.03$ \\
			$\mathcal{D}_{R}$ & $ -8.08 \pm 0.05$ & $ -8.53 \pm 0.05$ & $ -8.23 \pm 0.05$ & $ -7.84 \pm 0.05$ \\
			$\mathcal{D}_{170817+R}$ & $ 3.86 \pm 0.05$ & $ 4.97 \pm 0.05$ & $ 4.74 \pm 0.05$ & $ 4.91 \pm 0.05$ \\
			$\mathcal{D}_{\text{all}}$ & $ 21.87  \pm 0.05$ & $ 22.78 \pm 0.05$ & $ 22.32 \pm 0.05$ & $ 22.51 \pm 0.05$ \\
			\hline
	\end{tabular}}
	\caption{The log evidence $\ln p(\mathcal{D}|M)$ given the five different data sets when assuming each of the four models given the fitted Schechter luminosity function.}
	\label{tab:lnZFLF}
\end{table*}

\newpage

\section{Posteriors}

\begin{figure*}[!ht]
	\begin{center}
		\scalebox{1.7}{
        \includegraphics[width=0.45\textwidth]{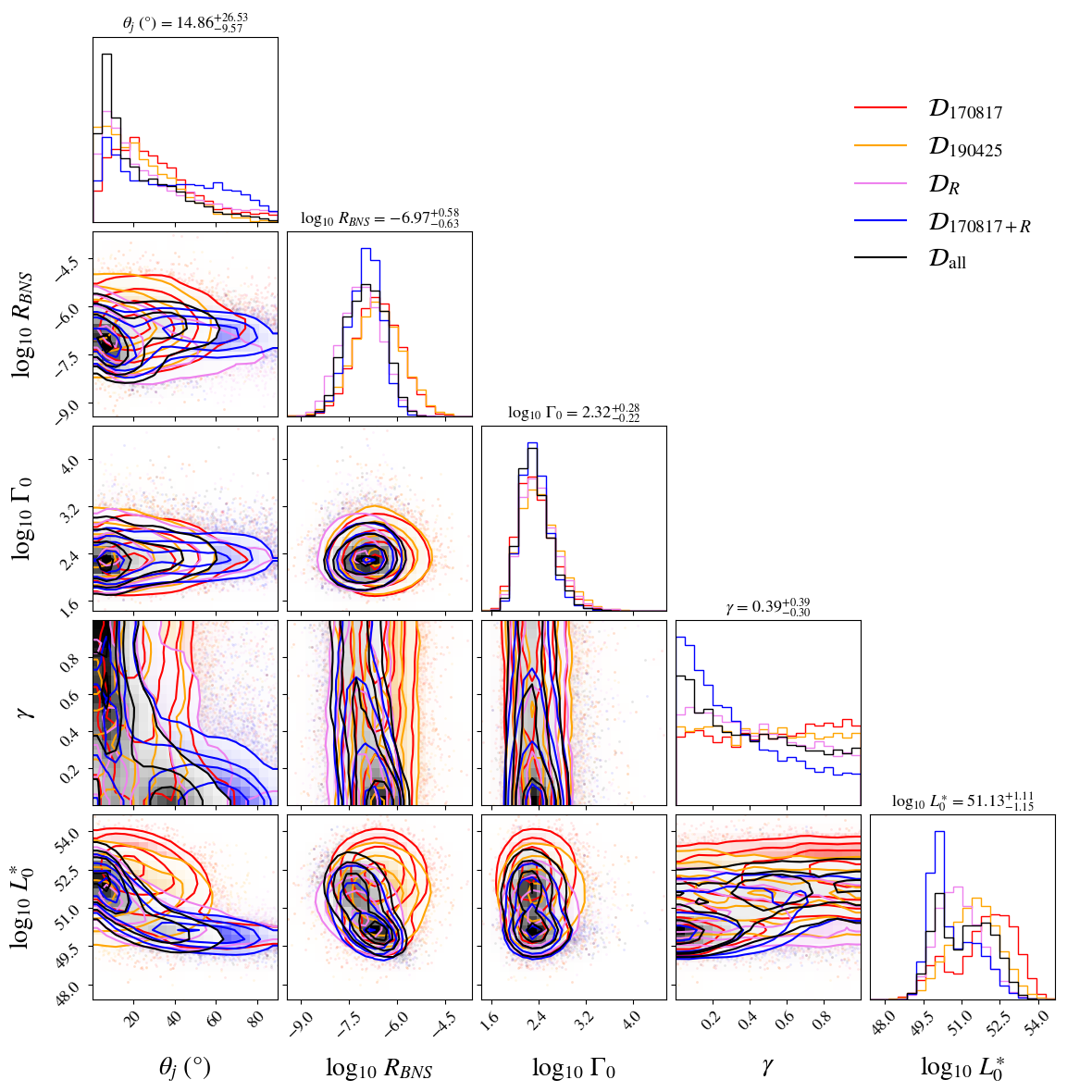}
		}
	\end{center}
	\caption{Parameter posterior for all data subsets when given the top-hat jet structure model.} 
	\label{fig:TH}
\end{figure*}

\begin{figure*}[!ht]
	\begin{center}
		\scalebox{1.7}{
        \includegraphics[width=0.45\textwidth]{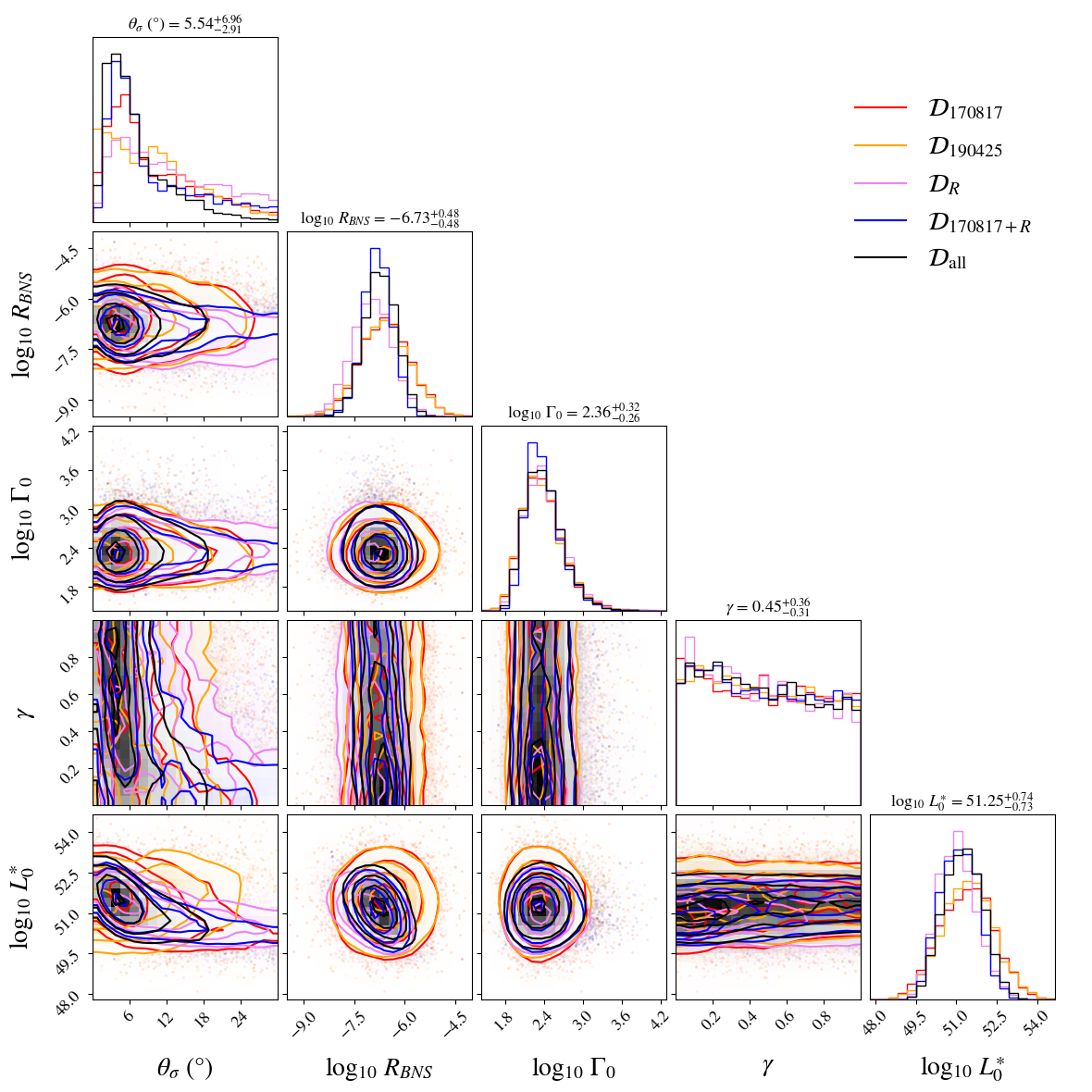}
		}
	\end{center}
	\caption{Parameter posterior for all data subsets when given the Gaussian jet structure model.} 
	\label{fig:GJ}
\end{figure*}

\begin{figure*}[!ht]
	\begin{center}
		\scalebox{1.7}{
        \includegraphics[width=0.45\textwidth]{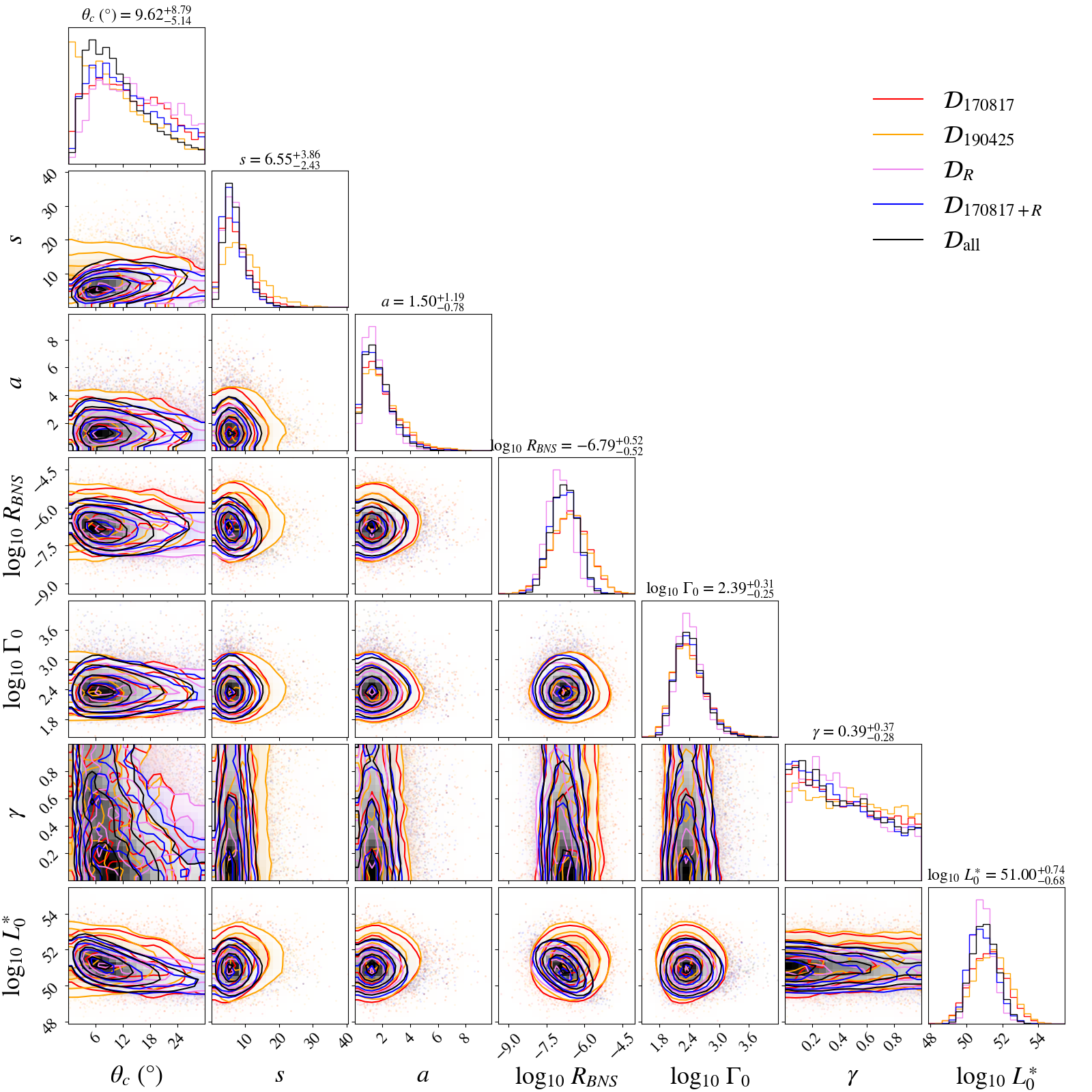}
		}
	\end{center}
	\caption{Parameter posterior for all data subsets when given the power-law jet structure model.} 
	\label{fig:PL}
\end{figure*}

\begin{figure*}[!ht]
	\begin{center}
		\scalebox{1.7}{
        \includegraphics[width=0.45\textwidth]{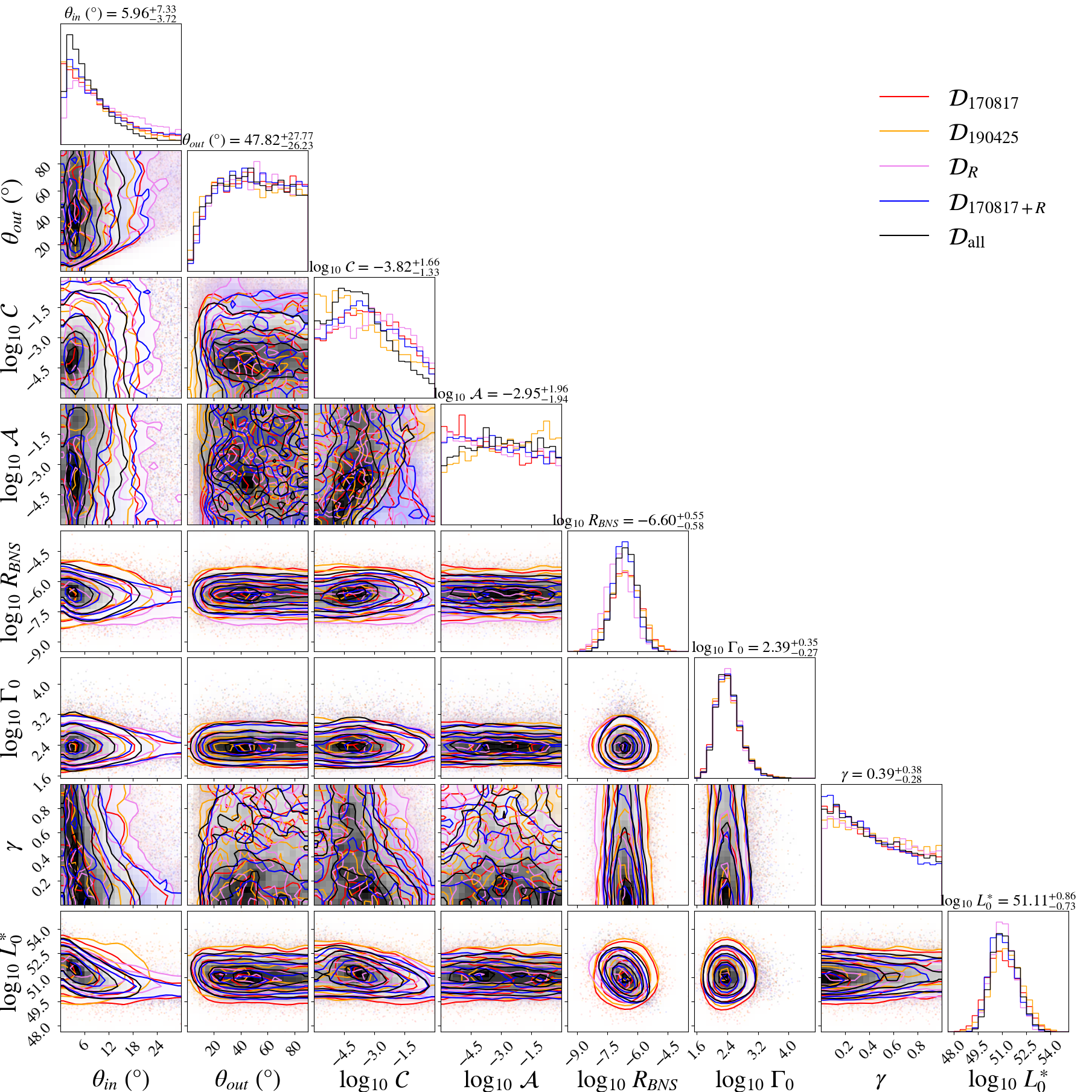}
		}
	\end{center}
	\caption{Parameter posterior for all data subsets when given the double Gaussian jet structure model.} 
	\label{fig:DG}
\end{figure*}



\end{document}